# Two novel methods for studying lowest frequency Earth oscillations before earthquakes

Randall D. Peters[1], and Michael A. Phillips[2]
[1]Physics Department, Mercer University, Macon, Georgia, and [2]Edward Pigot Seismic Observatory, Coonabarabran, Australia

## Abstract

Two novel methods, one that is experimental and the other comprising a pair of theoretical types (one component that is mathematically rigorous and the other that is of frequency domain computational type), are being used in concert to study the motion of the Earth at periods in the neighborhood of 1000 s to beyond 10,000 s. Corresponding to frequencies lower than what is routinely measured by commercial seismographs, our observations are yielding new insights into the mysteries of Earth dynamics responsible for earthquakes.

**Background-A simple yet relevant concept**

The first author of this article was trained in solid state physics, the field that came later to be known as condensed matter. His 1968 PhD dissertation was primarily concerned with the temperature dependence of 3rd order elastic constants of copper single crystals. Although defect states involving dislocations were not the principal focus of his research, nevertheless their influence was so important that some basic knowledge of their properties had to be learned. Those who specialize in materials science have even better understanding of crystalline defect structures, and nobody with knowledge of the subject should contest the validity of the left part of the following Fig. 1.

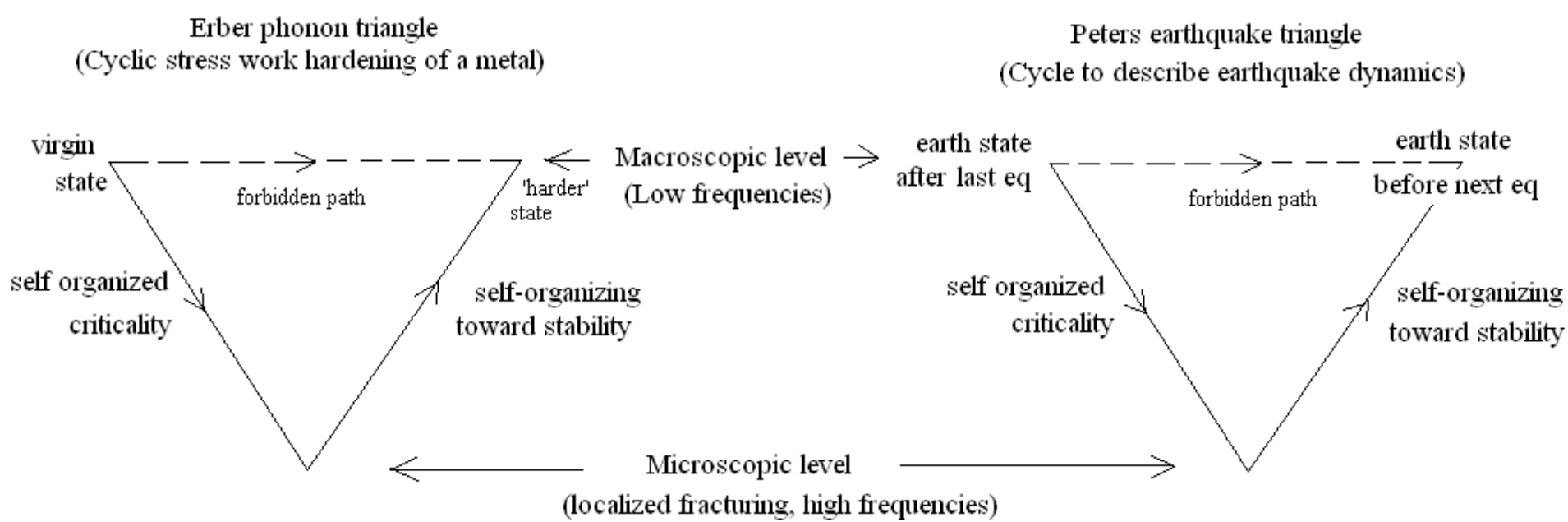

**Figure 1.** Illustration of the Erber Phonon triangle [1] (left) and the hypothesized similar earthquake triangle (right).

Cyclic stress applied to a metal sample causes it to follow the path shown in the left figure, going downward from the macroscopic state to the microscopic state. It is a path that involves avalanches and can be described by the theory of self organized criticality (SOC) developed by Bak et al [2]. The high frequency oscillations occurring at the microscopic level are of quantum type and called phonons. The downward left path is much better understood than the self organized stability path (upward right) [3] that is necessary before the sample undergoes the next transition that if continued indefinitely would result in sample failure (fracture, if the stress amplitude is sufficiently large).

Although the Earth at megameter scale and a metal sample at centimeter scale are radically different in size, one expects naturally that what is well known to materials science ought to have some applicability to the Earth, and therefore to the dynamics of earthquakes. Assuming the premise to be valid, then there is one obvious and unavoidable conclusion to be drawn. Central to the initiation of avalanches in a metal are the low frequency stresses applied to it. By analogy, we must pay attention to low-frequency forces that act on the Earth. The obvious one that should come to mind is that of the Moon/Sun tidal field. The horizontal (dashed line) of the phonon triangle is a forbidden path because none of the applied stresses are in the category of an enormous force that would be required to rupture the sample by means of a single application. Similarly the small tidal force disallows the horizontal (dotted) path of the earthquake triangle. Yet there must be consequence to the ever present 'massaging' of the Earth by it. And the multipli-fractured crust of the earth must undergo 'snap, cracle, pops' (and subsequent oscillations) like the shell of a hard boiled egg rolled between the hands. The global scale of the tidal force means that it causes frequencies of excitation that lie in the range of our planet's eigenmodes. So it is hypothesized that these eigenmode oscillations are the primary source of power with which to force the Earth down the SOC left path, toward an earthquake.

**Theoretical components of this study**

Support for the earthquake triangle shown in Fig. 1 derives also from (i) a publication that includes rigorous mathematical treatments [4], and a (ii) computational tool that is especially well suited to the investigation of temporal progressive changes in spectra [5]. Concerning [4], the mathematics of that article will prove challenging for the majority of individuals, including scientists. Only the following brief comments (central to the conslusions drawn from that work) need to be presently mentioned. The cm-sized sample that was studied showed a consistent 'critical slowing down of noise' before the appearance of a catastrophe. It was therefore natural for us to look for 'critical slowing down of Earth background noise before an earthquake'. The analysis methods of [4] to demonstrate that behavior were performed by the 2nd and 3rd authors, and could also be applied to the earthquake records presently studied. It was instead here chosen to demonstrate the behavior using a different method that yields similar conclusions, known as the "cumulative spectral power" (CSP) [5].

**Experimental components of this study**

All data considered for the generation of this article were provided by the 2nd author using the two channel VolksMeter (VM) that has been operative for several years at his facility, the Edward Pigot Seismic Observatory. A two-channel pendulum seismograph [6], the VM is especially well suited to the study of eigenmode (free) oscillations of the Earth. Its use of fully differential capacitive sensors to measure inertial mass displacement provides a significant advantage in that frequency realm over more conventional instruments that measure 'velocity'.

**Example earthquake record**

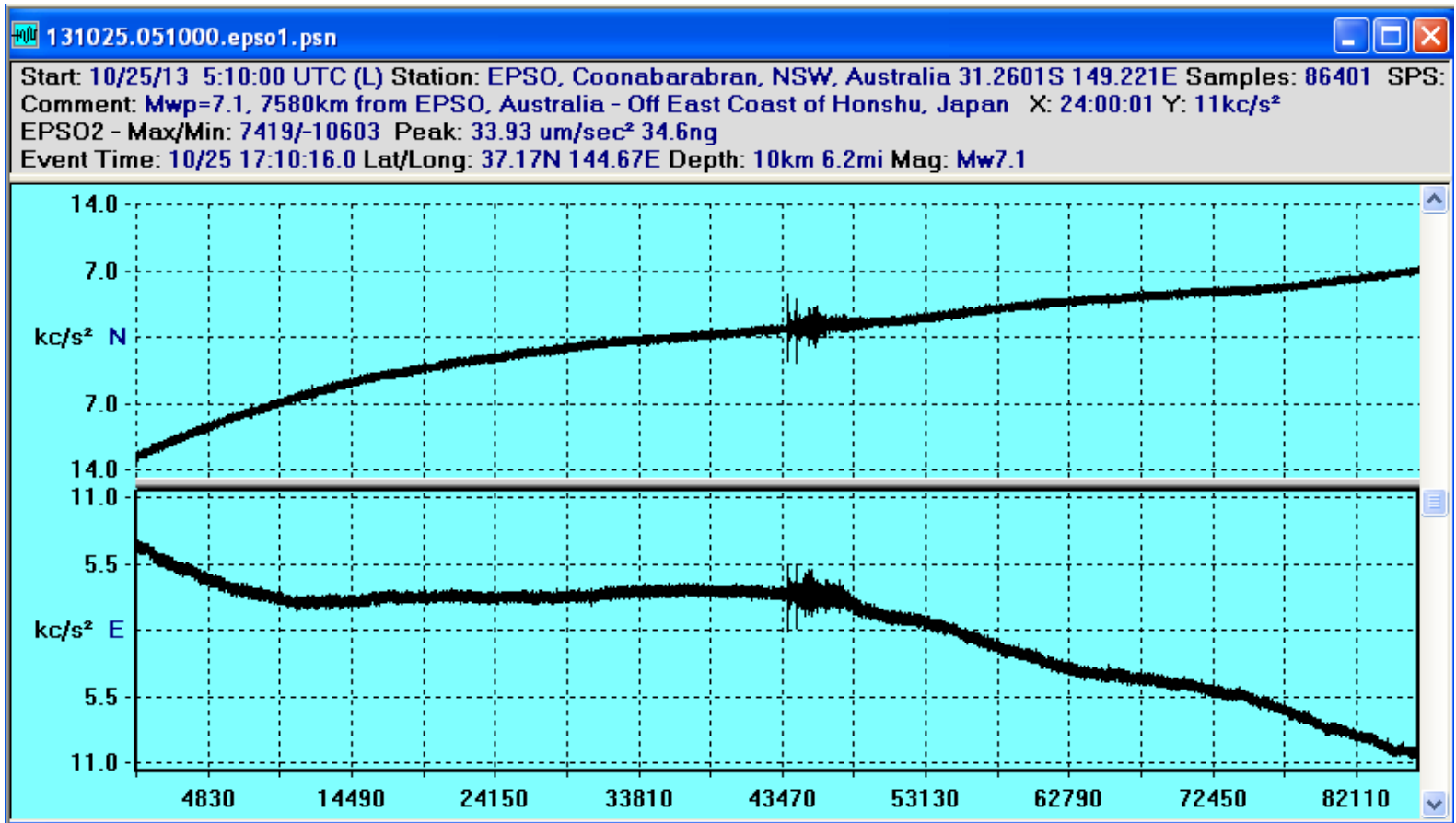

**Figure 2.** VM recordings of the M 7.1 earthquake off the east coast of Honshu, Japan 25 October 2013. Each record is approximately 24 hours duration, recording rate of one sample per second.

"Kink signatures' in the records of large earthquakes that were recorded by this instrument have been noted for many cases spanning an interval greater than 6 months. Only in the last week, however, have the powers of Mathematica been applied to the computation of their CSP's. Shown in Fig. 3 is an example of such, for the event pictured in Fig. 2.

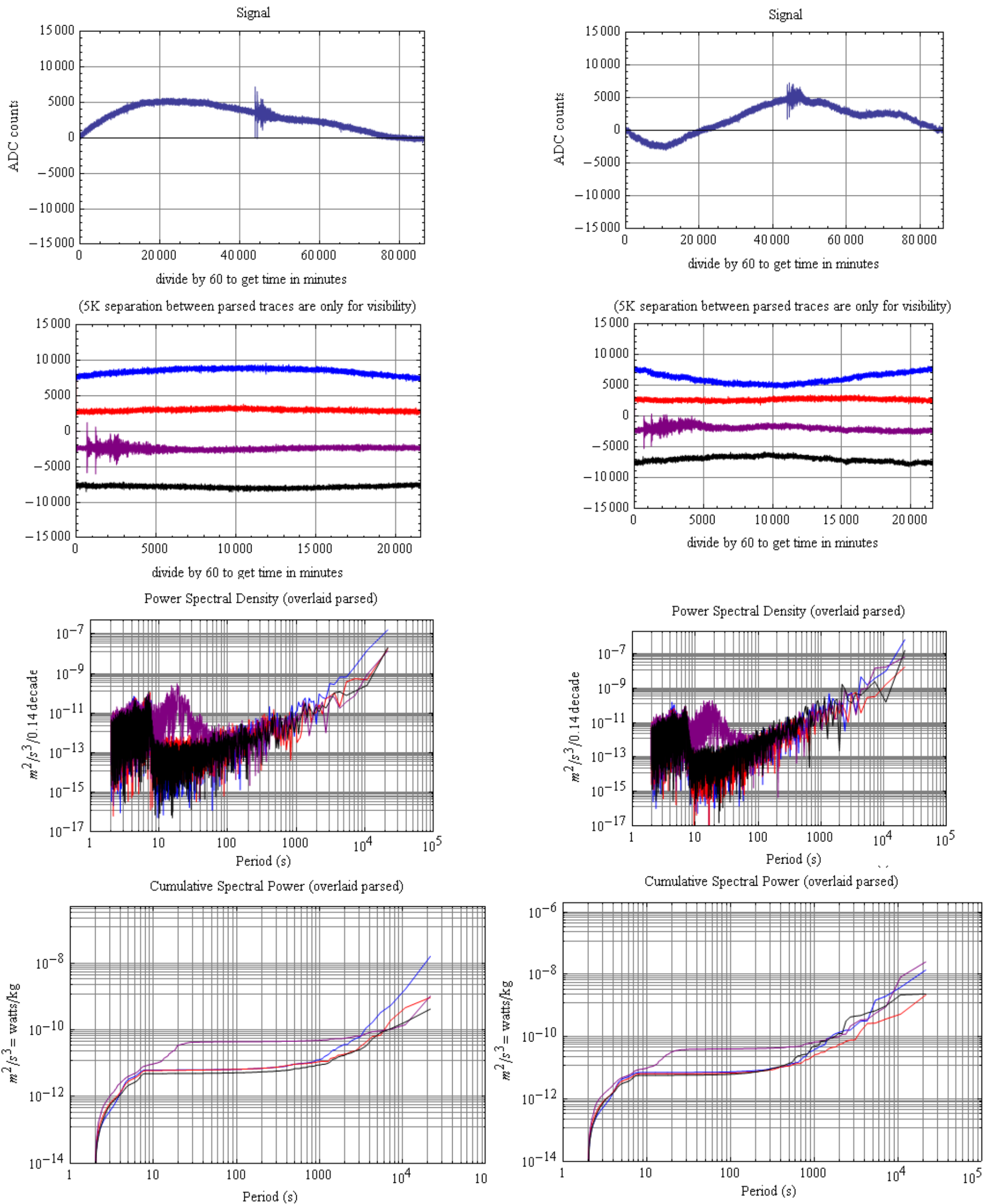

**Figure 3.** Graphics calculated using Mathematica analyses of the time records shown in Fig. 2. To demonstrate the temporal variations of the records over the one-day interval spanned, each was parsed into four equal (6 hr) pieces.

For the colored plots, the time sequence (for both temporal type and spectral type) progress from blue (start) through red, purple, and black (finish). To allow overlaid plots of the four parsed time parts and avoid interference with one another, adjacent pieces were separated by 5000 analog to digital counts of the VM output. The calibration constant for the instrument was assumed here to be 5.8 x 10$^8$ cts/rad; however, this parameter could be off by as much as a factor of 4 or 5. Nevertheless, the absolute values for the power spectral densities (PSD) and cumulative spectral powers (CSP) are not of great significance to the conclusions to be drawn from these graphs.

The third row of graphs in Fig. 3, demonstrate the difficulties of trying to draw frequency domain conclusions from the most common spectral representation type. The 'clutter' associated with FFT output (not noise, but just the nature of a spectrum) means that very many overlaid records 'wash out', so that any meaningful interpretation of their differences becomes quickly impossible. Although waterfall spectra have been used with some success to address this problem, the CSP is seen from the lowest row of graphs to be quite nice by comparison. Because it is obtained by integrating over the PSD, many of the bin to bin fluctuations of the PSD average in such a way as to yield a dramatically smoother curve. For those who as students learned that integration reduces noise and differentiation increases noise, this should come with little surprise. It also says something about the advantage of displacement measurements of inertial mass motion of a seismometer, as opposed to the conventional measurement of the derivative of displacement.

**Evidence for 'critical slowing down of noise before the earthquake'**

Consider the bottom row (CSP plots) of Fig. 3. In both the N/S (left) and E/W (right) channels of the VM output, we see in the purple curves the large increase in power (watts/kg) that occurred at the time the surface waves passed Coonabarabran. But notice the large increase in power for period beyond 1000 s in the N/S channel at the time of the first (blue) record. Although the blue power curve does not show such an increase in the E/W channel, there is in the subsequent (red) curve a signifcant shift of power toward longer periods. This is the hallmark of 'critical slowing down' before a catastrophe that was found to be reproducible in the research that yielded article [4]. It is as though the energy associated with north south eigenmode oscillations (time of blue curve) was large enough to exceed a critical threshold required for catastrophe, which did not happen without possible warning-by way of the red shift in the E/W record that happened before the event.

**Similar features in other EPSO records containing large earthquakes**

We do not here show other similar examples of what is seen in Fig. 3. Suffice it to say that there is compelling evidence from the majority of nearly a dozen records, to support even for earthquakes what was stated in [4], the abstract of which reads as follows:

Catastrophes of all kinds can be roughly defined as short-duration large-amplitude events following and followed by long periods of "ripening". Major earthquakes surely belong to the class of 'catastrophic' events. Because of the space-time scales involved, an experimental approach is often difficult, not to say impossible, however desirable it could be. Described in this article is a "laboratory" setup that yields data of a type that is amenable to theoretical methods of prediction. Observations are made of a critical slowing down in the noisy signal of a solder wire creeping under constant stress. This effect is shown to be a fair signal of the forthcoming catastrophe in both of two dynamical models. The first is an ästract" model in which a time dependent quantity drifts slowly but makes quick jumps from time to time. The second is a realistic physical model for the collective motion of dislocations (the Ananthakrishna set of equations for creep). Hope thus exists that similar changes in the response to noise could forewarn catastrophes in other situations, where such precursor effects should manifest early enough.